# Epitaxial Thin Films of a Chalcogenide Perovskite

*Mythili Surendran, Huandong Chen, Boyang Zhao, Arashdeep Singh Thind, Shantanu Singh, Thomas Orvis, Huan Zhao, Jae-Kyung Han, Han Htoon, Megumi Kawasaki, Rohan Mishra and Jayakanth Ravichandran[*]*

M. Surendran, T. Orvis
Mork Family Department of Chemical Engineering and Materials Science, and Core Center for Excellence in Nano Imaging, University of Southern California, 925 Bloom Walk, Los Angeles, CA 90089, USA

H. Chen, B. Zhao, S. Singh
Mork Family Department of Chemical Engineering and Materials Science, University of Southern California, 925 Bloom Walk, Los Angeles, CA 90089, USA

Dr. A.S. Thind
Institute of Materials Science & Engineering, Washington University in St. Louis, One Brookings Drive, St. Louis, MO 63130, USA

Dr. R. Mishra
Department of Mechanical Engineering & Materials Science, and Institute of Materials Science & Engineering, Washington University in St. Louis, One Brookings Drive, St. Louis, MO 63130, USA

Dr. H. Zhao, Dr. H. Htoon
Center for Integrated Nanotechnologies, Materials Physics and Applications Division, Los Alamos National Laboratory, NM 87545, USA

J.-K. Han, Dr. M. Kawasaki
School of Mechanical, Industrial & Manufacturing Engineering, Oregon State University, Corvallis, OR 97331, USA

Dr. J. Ravichandran
Mork Family Department of Chemical Engineering and Materials Science, and Ming Hsieh Department of Electrical and Computer Engineering, University of Southern California, 925 Bloom Walk, Los Angeles, CA 90089, USA
E-mail: j.ravichandran@usc.edu

**Abstract**

Chalcogenide perovskites have emerged as a new class of electronic materials, but fundamental properties and applications of chalcogenide perovskites remain limited by the lack of high quality epitaxial thin films. We report epitaxial thin film growth of $BaZrS_3$, a prototypical chalcogenide, by pulsed laser deposition. X-ray diffraction studies show that the films are strongly textured out of plane and have a clear in-plane epitaxial relationship with the substrate. Electron microscopy studies confirm the presence of epitaxy for the first few layers of the film at the interface, even though away from the interface the films are polycrystalline with a large number of extended defects suggesting the potential for further improvement in growth. X-Ray reflectivity and atomic force microscopy show smooth film surfaces and interfaces between the substrate and the film. The films show strong light absorption near the band edge and photoluminescence in the visible region. The photodetector devices show fast and efficient photo response with the highest ON/OFF ratio reported for $BaZrS_3$ films thus far. Our study opens up opportunities to realize epitaxial thin films, heterostructures, and superlattices of chalcogenide perovskites to probe fundamental physical phenomena and the resultant electronic and photonic device technologies.

Chalcogenide perovskites have recently[1-10] emerged as a new family of electronic materials that demonstrate exciting electronic and optical properties with promise for technological applications such as photovoltaics[4][8], infrared optics[11-13] and thermoelectrics[14, 15]. Chalcogenide perovskites such as $BaZrS_3$ (BZS) and related layered phases such as Ruddlesden Popper $Ba_3Zr_2S_7$ possess high absorption coefficients [8, 10, 16], long recombination lifetimes [17], and excellent thermal stability [18] rendering them as suitable candidates for solar energy conversion and optoelectronic applications [8, 10, 17]. Quasi-1D hexagonal chalcogenide perovskites such as $BaTiS_3$ and $Sr_{1+x}TiS_3$ exhibit optical anisotropy resulting in large birefringence and linear dichroism in the mid-wave and long-wave infrared region [11-13]. Over the years, advances in the synthesis of polycrystalline powders [6, 10, 19-21] and single crystals [22, 23] and theoretical methods have been critical to identify the desirable properties of chalcogenide perovskites. As we continue to understand the physical properties of chalcogenide perovskites, the next critical step is to develop the thin-film technology, especially epitaxial thin films. Recently, several groups have developed innovative, but indirect, approaches to achieve thin films of BZS, [5, 16, 24-26] but the direct layer-by layer growth of crystalline and/or epitaxial films remains a challenge. This step is pivotal to improve our understanding of their physical properties, explore emergent phenomena in heterostructures, strained thin films and superlattices, and lastly, enable broad technological applications that rely on thin-film technology.

Thin-film growth of chalcogenide perovskites presents three important hurdles: a large mismatch in the vapor pressure for the cations and chalcogens, corrosive and reactive nature of most chalcogen precursors, and the propensity to oxidize easily in the presence of oxygen at high temperatures. The former two may lead to strict limits on the sticking of films to the surface of the substrate, temperature range over which the growth is carried out, and interface reactions with the substrate, and the latter severely constrains the growth window. Recently, multiple groups have devised strategies to overcome these limitations. Polycrystalline thin films of BZS were achieved by sulfurization of $BaZrO_3$ films at high temperatures of about 1050°C using $CS_2$ as the sulfurizing agent [5, 16]. These films possessed high carrier densities, presumably arising from a large number of shallow donor point defects. BZS thin films were also synthesized by reactive sputtering at room temperature followed by a rapid thermal crystallization at 800-1000°C, resulting in polycrystalline films with secondary phases such as S-rich phases [24]. These reports clearly emphasize the difficulties in achieving large grain growth and stoichiometric composition for BZS in the thin film form. Several other synthesis routes for BZS thin films implemented thereafter such as sulfurization of $BaZrO_3$ using $Ar-H_2S$ gas [25], low temperature film deposition of amorphous BZS followed by a $CS_2$ annealing step [26] and solution processed colloidal BZS nanocrystals [27], yielded polycrystalline films with no preferred orientation. These studies highlight the synthetic difficulties in achieving high quality thin films of chalcogenide perovskites such as BZS, especially controlling their stoichiometry, crystallinity and texture. These polycrystalline films show promising electrical and optical characteristics despite the potential contaminants introduced by the two-step processes. More importantly, any two-step process involving deposition and annealing severely limits the ability to prepare multilayers and heterostructures needed for a broad range of applications. Hence, direct epitaxial growth is critical to not only disentangle the effects of extended structural defects such as grain boundaries and voids, but also expand the breadth of artificial material structures one can envision and grow at will.

In this article, we report the growth of epitaxial BZS thin films by pulsed laser deposition (PLD) at relatively low temperatures of 700 - 850°C, a critical step to enable the vision outlined above for chalcogenide perovskite heterostructures and superlattices. BZS thin films were grown from a phase pure BZS target in a background gas mixture of hydrogen sulfide and argon (Ar-$H_2$S) on perovskite oxide substrates such as $LaAlO_3$ (LAO) and $SrLaAlO_4$ (SLAO). The dense, polycrystalline target was prepared by sintering BZS powders by high pressure torsion (HPT) method [28, 29]. This method allows room temperature densification of BZS pellets, as high temperature sintering methods lead to nonstoichiometric compositions due to the propensity for loss of sulfur at elevated temperatures. The films grown at lower temperatures of ~400°C were amorphous, while epitaxial films were grown at 700-850°C [see Supporting Information for more details]. Films grown at all temperatures were dark in color suggesting strong absorption in the visible spectrum. All the characterization reported in this study were performed on as grown films without any further annealing.

**Figure 1**a shows a schematic illustration of BZS films grown on SLAO substrate. BZS has a lattice parameter, $a_{pc}$ = 4.968 Å ($a_{pc}$ is the pseudo-cubic lattice parameter derived from an orthorhombic

unit cell) and has an in-plane lattice mismatch of 8% with LAO substrate (pseudo-cubic lattice parameter = 3.79 Å) and 6.4% with the SLAO substrate ($a = b$ = 3.7564 Å, $c$ = 12.636 Å). Figure 1b shows a representative high resolution out-of-plane 2$\theta$-$\theta$ X-ray diffraction (XRD) scan of BZS films grown on SLAO and LAO substrates at 750°C. We observe only 202 reflection of BZS in these scans, which indicate a strong out-of-plane texture of the BZS films. Although the *d*-spacing for 040 and 202 reflections are very similar, STEM studies discussion below provides the basis for this out-of-plane orientation assignment, as opposed to the expected long axis orientation out-of-plane. Presumably the 101, 303 and 404 reflections were not observed due to the weak structure factor compared to the 202 reflection. BZS film grown on SLAO substrate showed a stronger out of plane texture compared to that on LAO substrate, likely due to a smaller lattice mismatch. The rocking curve of the 202 peak had a full-width half maximum (FWHM) of about 4º, as shown in the figure S3 of Supporting Information, indicating that the crystalline quality of the film can be further improved.

The in-plane epitaxial relationship between the film and the substrate was determined by pole figure analysis. We performed off-axis $\phi$-scans along BZS 121, LAO 103 and SLAO 103. Figure 1c shows four peaks each for BZS 121 and SLAO 103 separated by 90 degrees and aligned to each other. We did not observe any 45° offset reflections, as would have been expected if there were a 45° rotated alignment between BZS and SLAO. The expected 45º rotated cube on cube epitaxy was not observed in the $\phi$-scans likely due to the large lattice mismatch. Thus, the epitaxial relationship between the film and substrate is given as (101) [010] BZS // (001) [100] SLAO. A similar epitaxial relationship was found for the BZS film and the LAO substrate ((101) [010] BZS // (001) [100] LAO) (assuming a pseudo cubic unit cell for LAO), corresponding to an alignment of pseudo cubic edges of the film and the substrate. The measured lattice parameter from 2$\theta$-$\theta$

XRD scans implied that the epitaxial films were fully relaxed. The thicknesses of the films were measured by X-ray reflectivity (XRR) as shown in Figure 1d. The slow decay of the reflected X-ray intensity and the presence of Kiessig fringes indicate that the film has very low surface and interface roughness. AFM topography images further confirmed that the film surfaces are smooth with a sub-nanometer scale roughness for all the grown films. A representative case is shown in the inset of Figure 1d, where we observed a root mean squared (RMS) roughness of ~0.55 nm for 100 nm thick BZS films grown on LAO substrates.

To investigate the quality and structure of the epitaxial BZS film grown on SLAO substrates at the atomic level, we have performed scanning transmission electron microscopy (STEM). **Figure 2a** shows a wide field-of-view high-angle annular dark field (HAADF) image of the ~ 130 nm thick BZS film on SLAO substrate. As shown in Figure 2b, the BZS film is polycrystalline and contains extended defects such as grain boundaries and Ruddlesden-Popper planar faults that are commonly observed in perovskites [30, 31]. Figure 2c shows fast Fourier transform (FFT) patterns for the regions highlighted (red and green boxes in Figure 2b) indicating the changes in the crystallographic orientation of BZS film. Using the FFT spots highlighted in Figure 2c, we have generated the inverse FFT pattern for the region highlighted (white box) in Figure 2b. The inverse FFT pattern shown in Figure 2d clearly shows the presence of Ruddlesden-Popper planar faults and misfit dislocations at the BZS/SLAO interface.

Figure 2e shows an atomic resolution HAADF image of the BZS/SLAO interface, which confirms the presence of epitaxial relationship between the film and the substrate as (101) [001] BZS // (001) [100] SLAO for this region. We have further performed electron energy loss spectroscopy (EELS) to map the distribution of elements. Figure 2f shows a HAADF image, where EELS data was acquired for the region highlighted with a white box. Figure 2g shows the elemental maps for Ba $M$ edge, Zr $L$ edge, S $K$ edge, Sr $L$ edge, La $M$ edge, Al $K$ edge and O $K$ edge. From the Sr, La, Al and O maps and the HAADF images, it is clear that the substrate is terminated with LAO layers with a perovskite structure such that the epitaxial relationship is given by (101) [001] BZS // (001) [100] LAO. Figure 2h shows extracted EEL spectra for all the elements for three different regions (marked as red, blue and black regions in Figure 2f). Furthermore, the elemental maps for Ba, Zr and S reveal a uniform chemical distribution throughout the film.

Optical characterization of BZS thin films was carried out by room temperature photoluminescence (PL) measurements and UV-Visible transmission-reflection spectroscopy. Past experimental studies reported band gap ranging from 1.81 – 1.94 eV, including our report on the polycrystalline powder samples [10]. **Figure 3**a displays the Tauc plot extracted from UV-visible transmission-reflection measurements indicating an absorption edge close to 1.94 eV. The slightly higher band gap value has been attributed to the presence of oxygen in the BZS films [25], but other reports, presumably without significant oxidation have also reported a band gap of ~ 1.94 eV [8]. Further careful studies are needed to resolve this discrepancy in the band gap of BZS reported in the literature. High quality thin films allow determination of absorption coefficients from transmission and reflection measurements. BZS thin films have a high absorption coefficient ($\alpha$ >10$^5$ cm$^{-1}$) right near the onset of absorption near the band edge in agreement with the past reports of ellipsometry and UV-visible spectroscopy studies on ceramic pellets and polycrystalline thin films [5, 8, 16]. The PL spectrum shows a peak centered at 1.85 eV with a full width half maximum (FWHM) of 250 meV. The long tail of the PL curve at the lower photon energies indicates the presence of high density of defects close to the band edge. This behavior is consistent with our past observations in single crystals of BaZrS$_3$[32]. We also performed time-resolved photoluminescence (TRPL) measurements at room temperature to study the excited state dynamics. A biexponential fit to the data yields the time constants $\tau_1$ = 1.634 ns and $\tau_2$ = 8.06 ns, as shown in Figure 3b. The recombination time is an order of magnitude shorter than the reported time for $Ba_3Zr_2S_7$ single crystals [17] and two orders of magnitude shorter than the reported value for $BaZrS_3$ polycrystalline thin films [16]. As there is a broad variation in the structure, composition and processing conditions of these three cases, further studies are needed to understand the role of defects and intrinsic material properties on the recombination times of these chalcogenide perovskites.

The I-V characteristics of a BZS photodetector device were obtained in dark and under 532 nm, 633 nm and 785 nm laser illuminations, and are shown in figure S6 in the Supporting Information. The dark current from the BZS film was around 33 nA and a photo current of about 9.1 $\mu$A was measured at a power density of 113 mWcm$^{-2}$. The ON/OFF ratio of the device was found to be 275 at an applied bias of 10V and is the highest among all the BZS thin film-based photodetectors reported to date [5, 16, 26]. Past studies showed that the dark current of the BZS photodetector [5, 16] was highly sensitive to the point defects such as sulfur vacancies incorporated during the processing of the film. Wei *et al.* reported an order of magnitude higher dark current for BZS films synthesized by high temperature sulfurization [16], and later demonstrated a large reduction in dark current for BZS films grown at low temperature followed by a $CS_2$ annealing step as shown in their recent work [26]. The BZS thin films in our study were grown at an intermediate temperature of 750°C in a low concentration of $H_2S$ (4.75 mole %) background gas, with no further annealing steps. This can lead to sulfur deficient films and therefore can contain a non-trivial concentration of sulfur vacancies. Despite this, the dark current level was low with a higher ON/OFF ratio compared to previous reports on BZS thin film photodetectors.

The responsivity *(R)* of the photodetector as a function of wavelength at different applied voltages is shown in figure S7 of the Supporting Information. We observed a high responsivity of 17.5 mAW$^{-1}$ from an epitaxial BZS thin film detector under 532 nm excitation at an applied bias of 10 V. This value is higher compared to the recently reported 0.08 mAW$^{-1}$ responsivity for BZS films processed at lower temperatures [26] and is comparable to the 46.5 mAW$^{-1}$ responsivity at 5 V reported by Gupta *et al.* from CS$_2$ sulfurized BZS devices with five orders of magnitude higher dark current [5]. It can also be seen that the responsivity decreases as the excitation laser energy drops below band gap as expected from the absorption spectrum.

The photocurrent as a function of the illumination power can be fitted using a power–law relationship: I$_{ph}$ = AP$^\theta$ where *A* is a scaling constant, *P* is the incident light power and the exponent $\theta$ (0 < $\theta$ ≤ 1) to determine the mechanism behind the photocurrent generation in a photodetector. A value of $\theta$ = 1 implies that the increase in photocurrent is entirely due to photogenerated carriers, whereas $\theta$ < 1 indicates a complex process of carrier generation, recombination and charge trapping caused by trap states or defects [33-35]. **Figure 4**a shows the measured photocurrent at 10 V ($\lambda_{exc}$ = 532 nm) as a function of incident power density in a log-log representation along with the fitted curve. A relationship I$_{ph}$ ∝ P$^{0.71}$ was determined, with $\theta$ = 0.71 showing the effect of charge trapping probably due to the sulfur vacancies and/or oxygen substitutions on sulfur sites. For longer wavelengths, lower values of $\theta$ were observed due to inefficient light absorption and complex carrier generation and recombination mechanisms as shown in the Supporting Information.

The transient photo-response of the device at an excitation of 532 nm and applied bias of 10 V is shown in Figure 4b. The light source was periodically turned on and off at intervals of 30 s. A photocurrent of 2.32 $\mu$A with an ON/OFF ratio of 32 was measured at an incident light intensity of 20 mWcm$^{-2}$. The rise time (time to reach 90% of the photocurrent from dark current) and decay time (time needed to reach 10% of the photocurrent after switching off the illumination) was determined to be 0.72 s and 5.14 s respectively at 20 mWcm$^{-2}$ illumination intensity as shown in figure S6 of the Supporting Information. The slow decay time of the detector could be attributed to the presence of trap centers such as chalcogen vacancies [33].

To quantitively identify the optical performance of the epitaxial BZS detector, the photodetector characteristics were compared to the previously reported BZS thin films as indicated by figure 4c. Table 1 clearly illustrates the reported responsivity and ON/OFF ratios measured under similar conditions. The polycrystalline films from the literature reports showed no preferred texture, and the dark current was largely a function of processing conditions, especially temperature. Gupta *et al.* [5] reported the highest responsivity of 46.5 mAW$^{-1}$ from BZS films, but the dark current was extremely high due to elevated processing temperatures and hence, showed an ON/OFF ratio of ~ 1.5. On the other hand, Yu *et al.* [26] demonstrated that the BZS films annealed at lower temperatures greatly reduced the dark current showing an ON/OFF ratio of 80 and a responsivity of 0.08 mAW$^{-1}$. The epitaxial BZS films in this work showed the highest ON/OFF for BZS thin films (275 at 10 V) so far and high responsivity of 17.5 mAW$^{-1}$ with relatively low dark current. Photoresponse is extremely sensitive to the film quality in the bulk and on the surface. The enhanced photoresponse for epitaxial films can be attributed to the relatively high crystalline structure with lower defect density compared to recent reports. Thin film growth in higher H$_2$S concentrations or post growth annealing in sulfur rich conditions may shed light on the role of sulfur vacancies and/or oxygen substitutions in realizing high quality epitaxial thin films and high-performance optoelectronic devices. Our report opens up new avenues to accelerate the fundamental understanding and applications of chalcogenide perovskites.

**Conclusion**

In summary, we have demonstrated, for the first time, epitaxial growth of chalcogenide perovskite thin films by pulsed laser deposition on oxide substrates at temperatures as low as 700°C. Structural and chemical analysis showed clear epitaxial relationship between the film and the substrate with sharp substrate-film interface, albeit the films were polycrystalline with the presence of several extended defects. Optical characterization showed band gap close to 1.94 eV with strong light absorption near the band edge. The photodetectors fabricated showed high responsivity and fast response, with the highest ON/OFF ratio reported for BZS films thus far. This study opens up the opportunity to study the fundamental properties and emergent phenomena in chalcogenide perovskite heterostructures, and perovskite oxide/chalcogenide hybrid heterostructures. The realization of epitaxial thin film chalcogenide perovskites is a key step towards unlocking the promise for next generation electronic and photonic applications.

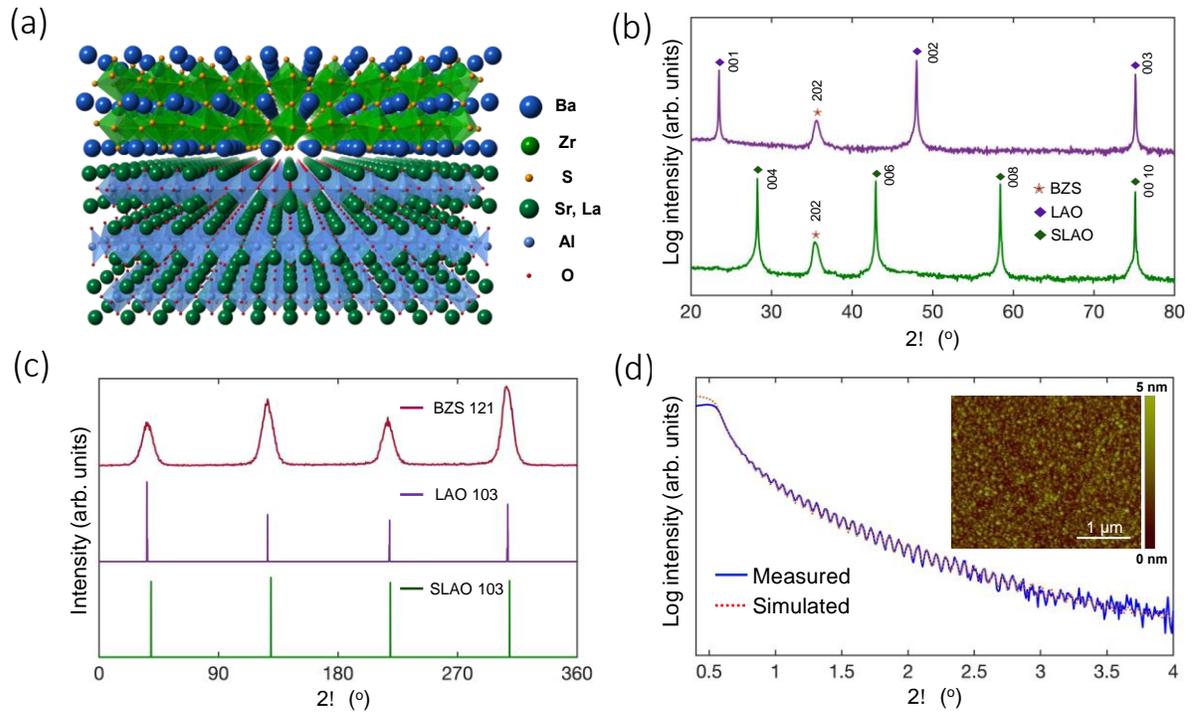

**Figure 1**. (a) Schematic of the crystal structure of the BZS / SLAO heterostructure, (b) High resolution 2$\theta$-$\theta$ XRD pattern of a representative BZS film on SLAO and LAO substrates, (c) Off-axis ϕ scans of BZS 121, LAO 103 and SLAO 103, (d) Measured (solid line) and simulated (dashed line) XRR curves of 100 nm BZS thin film on SLAO substrate. AFM image of as grown BZS film surface is shown in the inset.

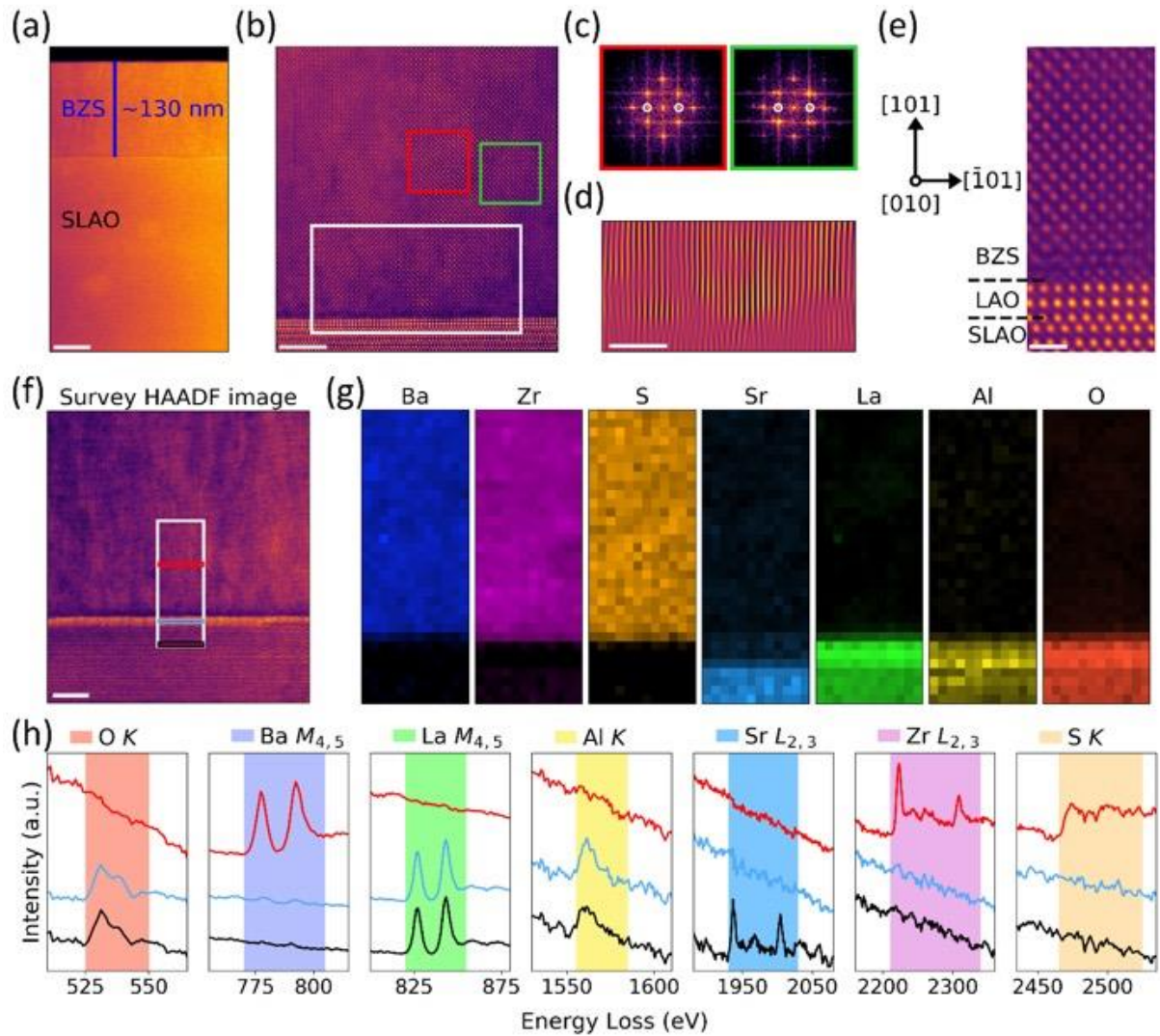

**Figure 2.** (a) Wide field-of-view HAADF image showing BZS film grown on an SLAO substrate. (b) HAADF image showing the polycrystalline nature of the BZS film. (c) FFT patterns indicating the changes in the crystallographic orientations for regions highlighted as red and green boxes in (b). (d) Inverse FFT for the region highlighted as white box in (b). The spots chosen for generating the inverse FFT are highlighted in (c). (e) Atomic resolution HAADF images showing the BZS/SLAO interface, with orientation of the BZS film for this region. (f) Survey HAADF image, where white box indicates the region from which the EELS data was acquired. (g) Elemental maps for Ba $M$, Zr $L$, S $K$, Sr $L$, La $M$, Al $K$ and O $K$ edges, for the region highlighted as white box in (f). Each elemental map is normalized within itself. (h) Extracted EEL spectra for all the elements, where the color of each spectrum corresponds to the region of same color highlighted as boxes in (d). Scale bars correspond to 50 nm for (a), 5 nm for (b) and (d), 1 nm for (e) and 10 nm for (f).

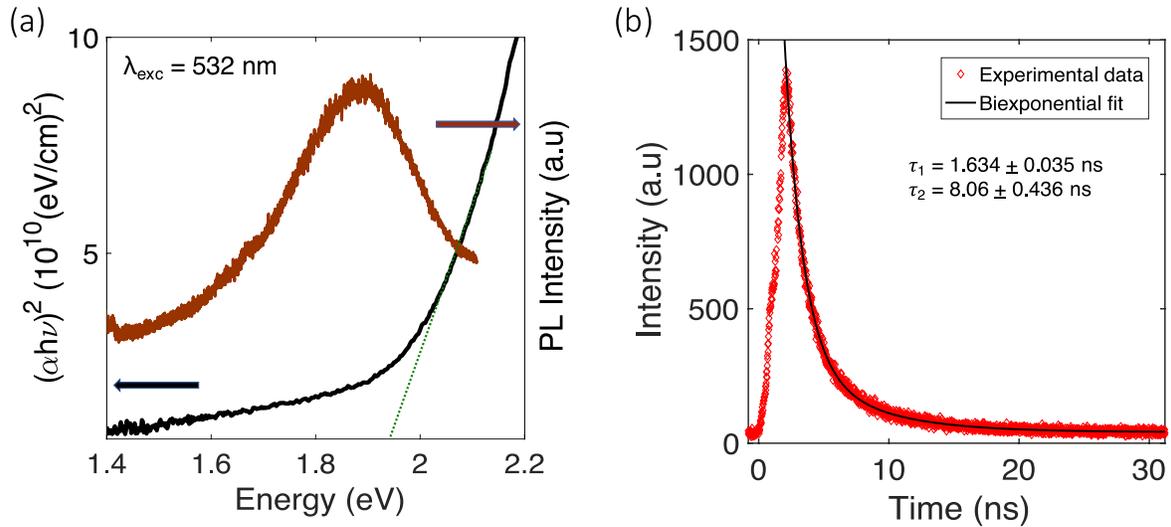

**Figure 3.** (a) Tauc plot extracted from UV-Visible transmission-reflection spectroscopy and PL spectrum of a 100 nm BZS film, (b) The TRPL decay profile of the emission peak (red). The black line shows a biexponential fit. We extracted the fast and slow time constants as $\tau_1$= 1.634 ns and $\tau_2$ = 8.06 ns respectively.

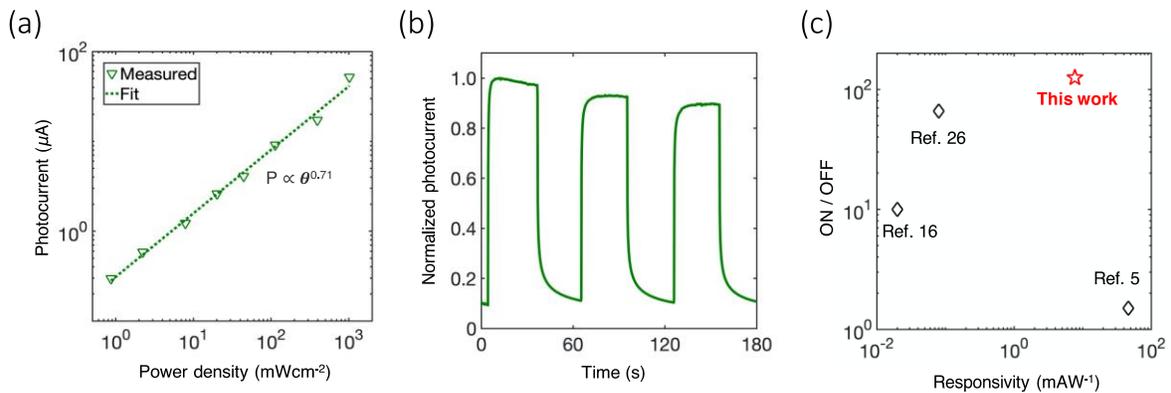

**Figure 4.** (a) Measured photocurrent and the corresponding power law fit (dotted line) as a function of incident power density under 532 nm excitation, (b) Measured transient photo response on-off characteristics with a periodically switched 532 nm laser source, (c) ON/OFF ratio and responsivity of epitaxial BZS films in comparison with previously reported values for BZS thin films.

**Table 1.** (Reported values of responsivity and ON/OFF ratio for BZS thin film photodetector devices)

| Responsivity [mAW$^{-1}$] (Bias applied) | ON/OFF ratio (Bias applied) | Wavelength [nm] | Reference |
|---|---|---|---|
| 0.02 (5V) | 20 (2V) | 532 | Ref. 16 |
| 46.5 (5V) | 1.5 (5V) | 405 | Ref. 5 |
| 0.08 (5V) | 80 (10V) | 532 | Ref. 26 |
| 7.7 (5V) | 125 (5V) | 532 | This work |
| 17.5 (10V) | 275 (10V) | | |

**Experimental Section**

*Thin film growth*: The BZS thin films were grown by PLD using a 248 nm KrF excimer laser in a hydrogen sulfide compatible vacuum chamber specifically designed for the growth of chalcogenide perovskites. Single crystal perovskite oxide (Crystec GmbH) substrates such as LAO and SLAO were pretreated by annealing at 1100°C for 3 h in 100 sccm $O_2$ and subsequently cleaned in acetone and IPA prior to deposition. The chamber was evacuated to a base pressure of $10^{-8}$ mbar and then backfilled with high purity argon gas. The substrate was heated up to the growth temperature in an argon partial pressure of 10 mTorr. A dense polycrystalline 1-inch BZS pellet synthesized by high pressure torsion process under 6 GPa was used as the target, which was preablated before growth. Argon – $H_2S$ (4.75 mole %) gas mixture was introduced into the chamber right before the growth to obtain an optimum growth pressure of 5 mTorr. The fluence was fixed at 2.0 J/cm$^2$ and the target- substrate distance used was 75 mm. The films were cooled postgrowth at a rate of 5°C/min.

*Structural and surface characterization*: The high resolution out of plane XRD scans and the off-axis scans for pole figure analysis were carried out on a Bruker D8 Advance diffractometer using a Ge (004) two bounce monochromator with Cu K$\alpha_1$($\lambda$ = 1.5406 Å) radiation at room temperature. X-ray reflectivity measurements were done on the same diffractometer in a parallel beam geometry using a Göbel (parabolic) mirror set up. Atomic force microscopy (AFM) was performed on Bruker di Multimode V atomic force microscope in non-contact mode to obtain the surface morphology and roughness.

*Electron Microscopy*: STEM experiments were carried out using the aberration-corrected Nion UltraSTEM$^{TM}$ 200 (operated at 200kV) microscope at Oak Ridge National Laboratory, which is equipped with a fifth-order aberration corrector and a cold field emission electron gun. EEL spectroscopy was carried out using a dual-range Gatan Enfinium spectrometer. A collection semi-angle of 33 mrad and an energy dispersion of 1 eV per channel was used to acquire EELS datasets. Pixel dwell time of 0.1 seconds and 2 seconds were used for low and high-loss EELS data acquisition. We have performed principal component analysis (PCA) for the EEL spectrum data to improve the signal-to-noise ratio. A power law was used to model the background signal prior to the core-loss signal for each element.

The cross-section sample for STEM characterization was prepared by parallel polishing using a MultiPrepTM system followed by low angle ion milling. The sample was parallel polished to a thickness of about ~20 μm. The ion milling was carried out at 5 kV followed by final polishing at 0.5 kV, using Fischione model 1010 low angle ion milling and polishing system. The cross-section sample was baked at 160 ˚C under vacuum for ~8 hours prior to the STEM experiments.

*Optical Spectroscopy:* The room temperature PL spectroscopy measurements were performed in a Renishaw inVia confocal Raman Microscope using a 532 nm diode laser through a 100X objective with a numerical aperture of 0.1. The excitation power used was ~1 mW. Reflectance and transmission measurements were conducted in a UV-Vis-NIR spectrometer (Lambda 950, PerkinElmer) from 400-1100 nm, and the absorption coefficient ($\alpha$) was determined using Beer-Lambert law $\alpha = -(1/d) \log(T/(1-R)^2)$ where $d$ is the film thickness, $T$ and $R$ are transmittance and reflectance, respectively. For time-resolved decay measurements, PL was confocally excited by a 60 ps, 405 nm laser pulses at 5 MHz repetition rate through a 50X objective with a numerical aperture of 0.7. The excitation power used was ~ 100 $\mu$W. The collected PL was spectrally filtered using a combination of 620 nm long-pass filter and a 680 nm short-pass filter and detected by a pair of fast avalanche photodiodes with 16 ps time resolution. The measurements were carried out with a HydraHarp 400 Time-Correlated Single Photon Counting (TCSPC) system.

*Photodetector device fabrication and response measurements*: Planar photodetector devices were fabricated with finger geometry (~20 µm gap size and 20 µm finger width). Electrodes of Ti/Au=3/80 nm were deposited using standard photolithography and ebeam evaporation. I-V characterization and transient photocurrent measurements were conducted using a semiconductor parameter analyzer (Agilent 4156C). Photoexcitation was introduced by Renishaw inVia confocal Raman Microscope using 532 nm / 633 nm / 785 nm diode laser as excitation through a 10X objective with numerical aperture from 0.0005 to 1. The excitation power through the objective was measured by a calibrated Thorlabs PM100D Power and Energy Meter Console.

**Supporting Information**

Supporting Information is available from the Wiley Online Library or from the authors.

**Conflict of Interest**

The authors declare no conflict of interest.

**Acknowledgements**

This work was supported by the Army Research Office under Award No. W911NF-19-1-0137 and the USC Provost New Strategic Directions for Research Award. R.M. and A.S.T. acknowledge support from the National Science Foundation through grant number DMR-1806147. The work of HPT processing at Oregon State University was supported by the National Science Foundation of the United States under Grant No. DMR-1810343. TRPL experiments were conducted at the Center for Integrated Nanotechnologies, an Office of Science User Facility operated for the U.S. Department of Energy (DOE) Office of Science and supported by DOE BES, QIS Infrastructure Development Project "Deterministic Placement and Integration of Quantum Defects". The authors gratefully acknowledge the use of facilities at Dr. Stephen Cronin's Lab, John O'Brien Nanofabrication Laboratory and Core Center for Excellence in Nano Imaging at University of Southern California for the results reported in this manuscript. STEM sample preparation was conducted at the Center for Nanophase Materials Sciences at Oak Ridge National Laboratory (ORNL), which is a Department of Energy (DOE) Office of Science User Facility, through a user project (A.S.T. and R.M.). Microscopy work performed at ORNL was supported by the U.S. DOE, Office of Science, Basic Energy Sciences Materials Science and Engineering Division (BES-MSED). We acknowledge helpful discussions with Dr. Rafael Jaramillo of the Massachusetts Institute of Technology, whose research group recently achieved similar epitaxial BZS film growth using molecular beam epitaxy.

# Supporting Information

**Epitaxial Thin Films of a Chalcogenide Perovskite**

*Mythili Surendran, Huandong Chen, Boyang Zhao, Arashdeep Singh Thind, Shantanu Singh, Thomas Orvis, Huan Zhao, Jae-Kyung Han, Han Htoon, Megumi Kawasaki, Rohan Mishra and Jayakanth Ravichandran[*]*

**High Pressure Torsion sintered BZS target for PLD**

Stoichiometric quantities of barium sulfide powder (Alfa Aesar 99.7%), zirconium powder (STREM, 99.5%), sulfur pieces (Alfa Aesar 99.999%), and iodine pieces (Alfa Aesar 99.99%) were mixed in a glove box and sealed in a quartz tube and was held at 960°C for 150 h to obtain phase pure BZS powders. The samples were ground and pressed into a 1-inch pellet using a hydraulic cold press. The pellets were further sintered using high pressure torsion at room temperature where a torsional strain is introduced by simultaneous compressive force and sample rotation to achieve a density of over 90%. **Figure S1** shows the powder XRD scans of as-synthesized BZS powders and HPT sintered BZS target.

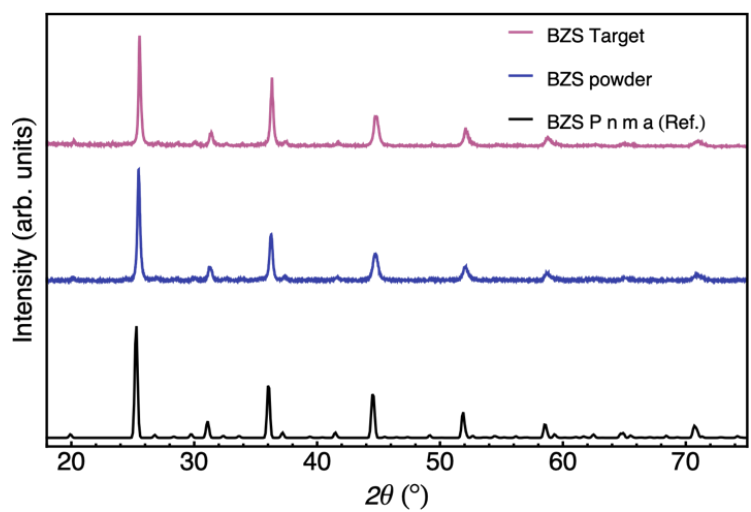

**Figure S1**. Powder XRD patterns of as synthesized BZS powders and torsional sintered BZS target along with BZS P*nma* reference

**Temperature dependence on texture**

BZS films grown at 750°C in Ar-H$_2$S showed the strongest texture in out of plane XRD scans. Films grown at temperatures below 700°C showed no texture and are most likely amorphous or poorly textured due to the low growth temperatures. XRD plot of a high resolution out of plane 2$\theta$-$\theta$ scans performed on 120 nm BZS films grown on LAO substrates at 650 – 850°C is shown in **Figure S2**. Above 800°C, an additional peak was observed at 37.5°, probably due to a partially strained film. A detailed discussion of strain effects and structural changes induced by lattice mismatch is beyond the scope of the current study and will be pursued in the future. No peak was observed for films grown at 650°C.

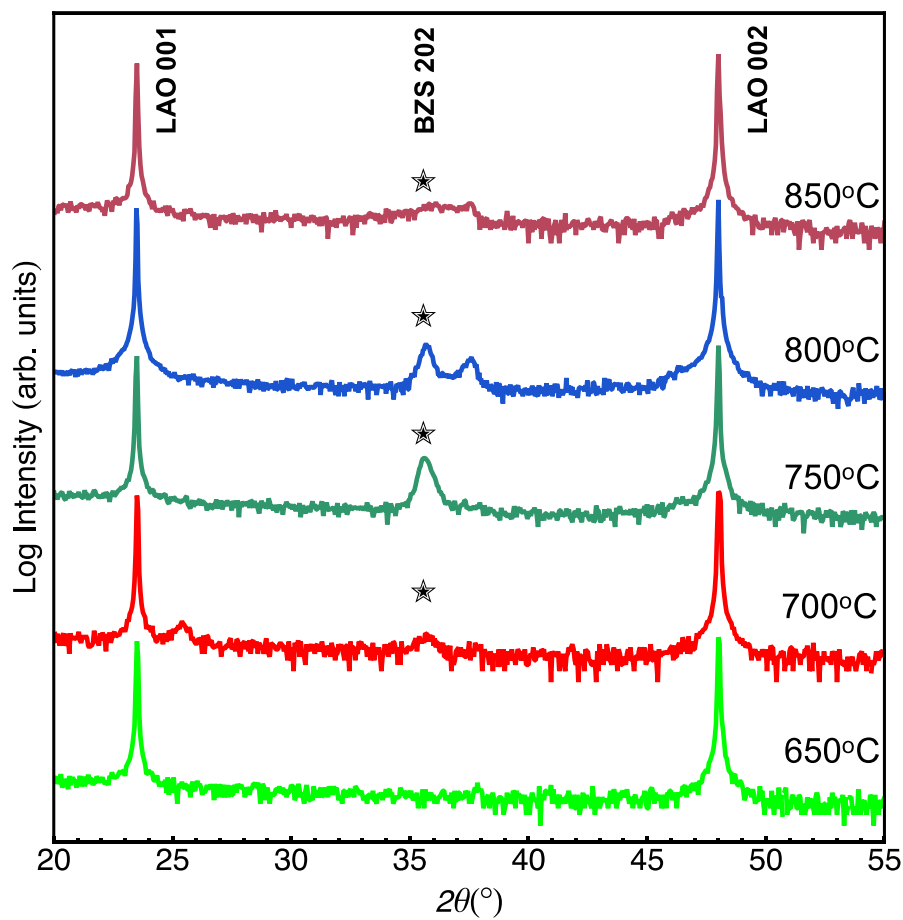

**Figure S2**. High resolution 2$\theta$-$\theta$ XRD pattern of BZS films on LAO substrate grown at 650-850°C in Ar-H$_2$S gas. Films grown at and above 700°C showed an out of plane texture.

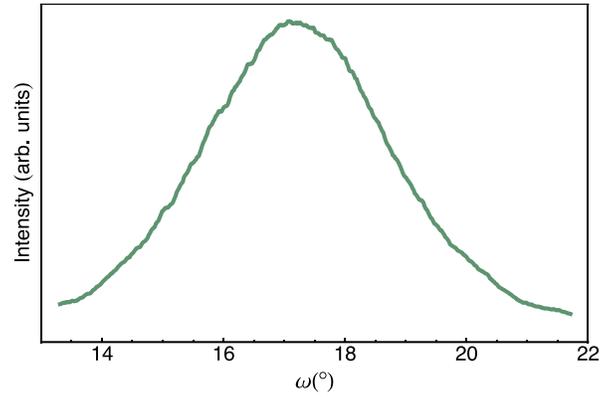

**Figure S3**. Rocking curve of 202 reflection of BZS film (FWHM ~ 4°) grown on LAO substrate at 750°C

**UV-Visible spectroscopy**

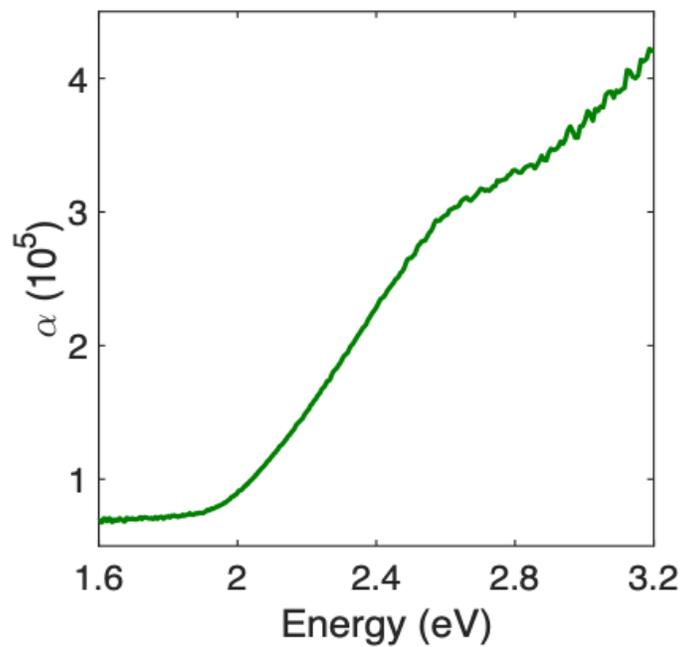

**Figure S4.** Absorption coefficient (in cm$^{-1}$) extracted from UV-visible transmission-reflection spectroscopy on a 100 nm BZS film.

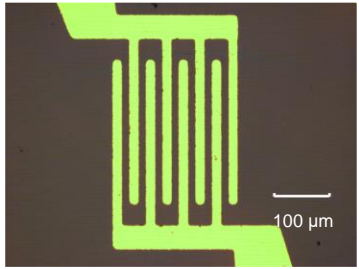

**Figure S5**. Optical image of a 100 nm thick BZS / SLAO photodetector device

**Photodetector measurements**

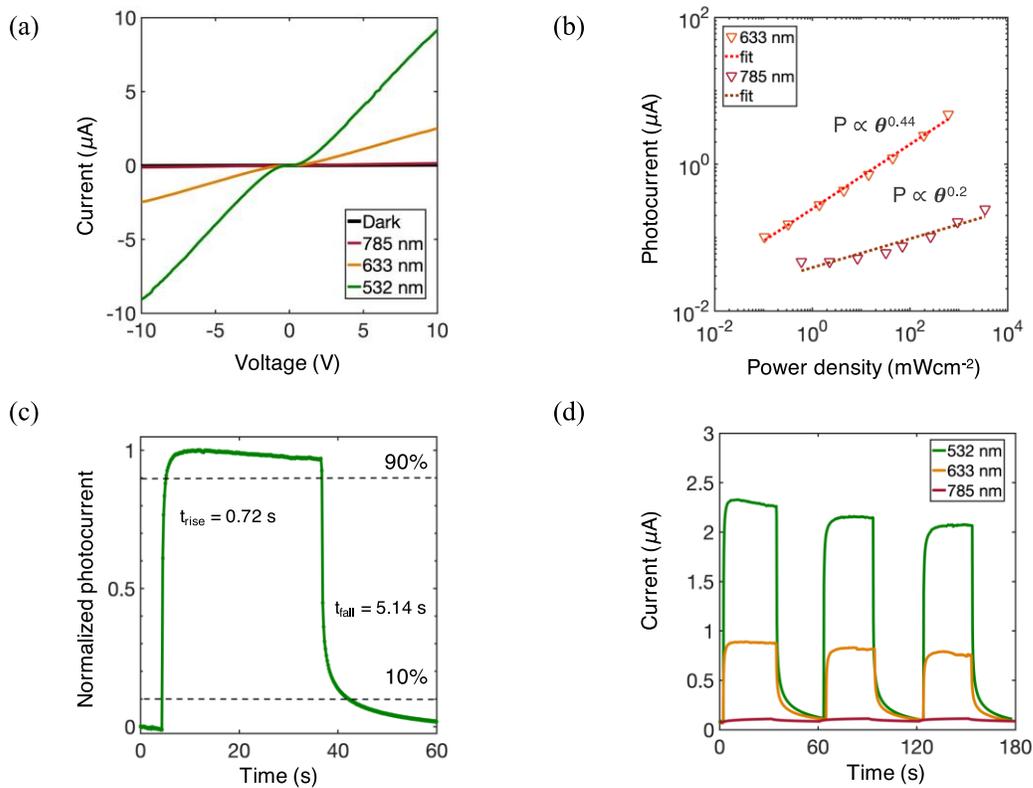

**Figure S6**. (a) I-V characteristics of a BZS photodetector device upon illumination with 532 nm, 633 nm and 785 nm laser sources, (b) Measured photocurrent and the corresponding power law fits (dotted line) as a function of incident power density for 633 nm and 785 nm excitations, (c) Normalized photocurrent vs. time plot showing a rise time of 0.72s and a fall time of 5.14s, (d) Transient photo response at 10V under different excitation wavelengths. 785 nm (below band gap) excitation showed very poor response.

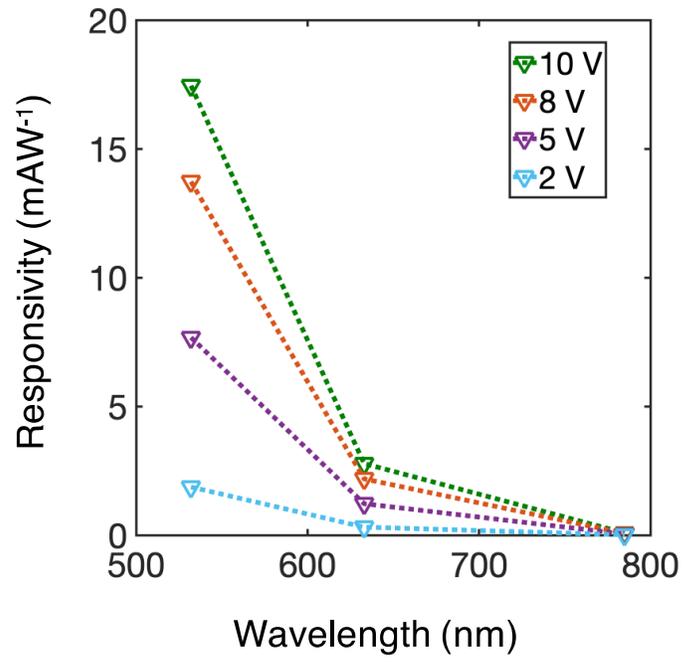

**Figure S7.** Spectral responsivity of the photodetector device at different applied voltages